\documentclass[letters,fleqn,usenatbib,useAMS]{mnras}
\usepackage[T1]{fontenc}
\usepackage{aecompl}
\usepackage{amsmath}
\usepackage{amssymb}
\usepackage{braket}
\usepackage{bm}
\usepackage{aas_macros}
\usepackage{ulem}
\usepackage{color}
\usepackage{graphicx}
\title{Magnetic reheating}

\author[S. Saga et al.]{
Shohei Saga,$^{1}$\thanks{shohei.saga@yukawa.kyoto-u.ac.jp}
Hiroyuki Tashiro,$^{2}$
and Shuichiro Yokoyama$^{3,4}$
\\
$^{1}$Yukawa Institute for Theoretical Physics, Kyoto University, Kyoto 606-8502, Japan\\
$^{2}$Department of Physics and Astrophysics, Nagoya University, Nagoya, 464-8602, Japan\\
$^{3}$Department of Physics, Rikkyo University, Tokyo 171-8501, Japan\\
$^{4}$Kavli IPMU (WPI), UTIAS, The University of Tokyo,
Kashiwa, Chiba 277-8583, Japan}
\begin{document}
\maketitle
\begin{abstract}
We provide a new bound on the amplitude of primordial magnetic fields (PMFs)
by using a novel mechanism, {\it magnetic reheating}.
The damping of the MHD fluid motions in a primordial plasma brings the
dissipation of the PMFs.
In the early Universe with $z \gtrsim 2 \times 10^6$, cosmic microwave
background (CMB)
photons are quickly thermalized with the dissipated energy and shift to a
different Planck distribution with a new temperature.
In other words, the PMF dissipation changes the baryon-photon
number ratio and we name such a process {\it magnetic
reheating}.
From the current baryon-photon number ratio
obtained from the BBN and CMB observations, we put a
strongest constraint on the PMFs on small scales which CMB
observations can not access, $B_{0} \lesssim 1.0 \;
\mu{\rm G}$ at the scales $10^{4} \; h{\rm Mpc}^{-1} < k < 10^{8} \; h{\rm
Mpc}^{-1}$. Moreover, when the PMF spectrum is given in a blue
power-law type, the magnetic reheating puts a quite strong constraint,
for example, $B_0 \lesssim 10^{-17} \;{\rm nG}$,
$10^{-23} \;{\rm nG}$, and $10^{-29} \;{\rm nG}$ at 1~comoving Mpc for
$n_{B}=1.0$, $2.0$, and $3.0$, respectively. This constraint would give an
impact on generation mechanisms of PMFs in the early Universe.
\end{abstract}

\begin{keywords}
cosmology: theory --- cosmology: cosmic background radiation --- cosmology: early Universe
\end{keywords}

%=========================================
\section{introduction}
%=========================================

Magnetic fields are ubiquitous in the universe and they have been observed in various astrophysical objects from planets and stars to galaxies and galaxy clusters~\citep{2012SSRv..166....1R,2012SSRv..166...37W}.
Recently there are some reports to suggest the existence of magnetic fields in the intergalactic medium~\citep{Neronov:1900zz,Tavecchio:2010mk,Essey:2010nd,Tashiro:2013ita,Chen:2014rsa}.
Primordial magnetic fields~(PMFs) are very attractive as the origin of observed magnetic fields on cosmological scales.
There are many proposals of the PMF generation mechanism in the early universe~\citep{Durrer:2013pga,Subramanian:2015lua}.
Therefore, observational constraints on PMFs can provide insights into the physics in the early universe.

PMFs can leave their signatures on many cosmological phenomena.
As examples, the requirement of successful Big Bang nucleosynthesis~(BBN) gives a constraint on the existence of PMFs at that time.
The recent detailed study about the BBN with PMFs provides the upper limit $B_0 <1.5~\mu$G where $B_0$ represents the present strength of PMFs~\citep{2012PhRvD..86f3003K}.
The measurements of cosmic microwave background~(CMB) anisotropies also yield constraints on PMFs, because PMFs can create the additional features in the temperature and polarization anisotropies of the CMB~\citep{1998PhRvL..81.3575S,2010PhRvD..81d3517S}.
Recent {\it Planck} data gave a limit on PMFs as~$B_0 \lesssim \mathcal{O}(1)~$nG around comoving $1$~Mpc, depending on the scale dependence of PMFs~\citep{2016A&A...594A..19P}.

Moreover, dissipation of PMFs in the early universe has also drawn attention to obtain a constraint on 
PMFs at smaller scales, compared with the scales which could be constrained from the CMB temperature and polarization anisotropies.
Before the epoch of recombination, PMFs induce the fluid motions in a photon-baryon plasma through the Lorentz force.
From the viewpoint of the energy transfer, this process can be considered as a conversion of the magnetic fields' energy to the fluid kinetic energy.
Since there arises a viscosity in the plasma due to the finite mean free path of photons, the induced motions on small scales are damped~\citep{Jedamzik:1996wp,Subramanian:1997gi}.
As a result, the magnetic fields' energy on small scales would be dissipated into the photon-baryon plasma.
If the dissipation happens after the redshift $z \sim 2\times 10^6$~(for recent reviews see~\citealt{2012MNRAS.419.1294C,Tashiro:2014pga}),
these dissipated energy can be observed as CMB distortions, that is, the spectral deviation of the CMB from the Planck distribution~\citep{Jedamzik:1999bm,Miyamoto:2013oua,2014JCAP...01..009K,Ganc:2014wia}.
Depending on the redshift of the energy injections, the generated CMB distortions are typically characterized by two parameters, the chemical potential $\mu$ and the $y$ Compton parameter.
Therefore, the measurement of CMB distortions allows us to understand the thermal history of the universe
from the redshift $z \sim 2\times 10^6$ to the epoch of recombination.
The current constraint on CMB distortions has been obtained by COBE/FIRAS~\citep{Fixsen:1996nj}.
According to the COBE/FIRAS constraint, the upper limit on the PMFs is set to $B_{0} < 30~{\rm nG}$ between comoving $400$~pc and $0.5$~Mpc ~\citep{Jedamzik:1999bm}.

In this Letter, we focus on the dissipation of the magnetic fields' energy before the CMB distortion era~($z \gtrsim z_{\mu} = 2.0\times 10^{6}$).
Before the CMB distortion era, any kind of the energy injections does not distort the energy spectrum of CMB photons.
This is because the interaction processes for the CMB thermalization, {\it i.e.,}~the Compton scattering, double Compton scattering, and bremsstrahlung have smaller time scales than the cosmological time scale.
In particular, the double Compton process keeps the distribution of CMB photons as the Planck distribution by adjusting the number of CMB photons, that is, the CMB temperature.
As a result, a {\it reheating} of the CMB photons may occur.

If we measure the overall history of the absolute CMB temperature
or the number density of photons before the CMB distortion era, we can
directly put a strong constraint on the amount of the injected energy.
However, the cosmological observations, e.g., the measurement of CMB anisotropies, tell us the information only around the recombination epoch\footnote{As we have mentioned, CMB distortions may tell us the information deeply before the recombination epoch.}.
In this Letter, we therefore use the alternative observable, that is, a baryon-photon ratio $\eta = n_{\rm b}/n_{\gamma}$.
Before the CMB distortion era, an energy injection into CMB photons can
increase the CMB photon number due to the double Compton process and bremsstrahlung, while the number of baryons does not change.
As a result, the baryon-photon number ratio decreases after the energy
injection occurs.
Since the light element production in the BBN is sensitive to $\eta$ at
the BBN epoch, the measurement of the light element amount
can observationally determine the value of $\eta$, $\eta_{\rm BBN}$, at the BBN epoch with some errors.
Independently, CMB observations can provide the value of $\eta$, $\eta_{\rm CMB}$, around the epoch of recombination with some errors.
If the energy injection to the CMB photons occurred after the BBN era
and before the CMB distortion era, the difference of the baryon-photon number ratios at between the BBN era
and CMB era can be reread in terms of the injected energy density of
photons $\Delta \rho_\gamma$ between these eras as
\begin{equation}
\frac{\eta_{\rm CMB}} {\eta_{\rm BBN}}
 = 1 - \frac{3}{4}\frac{\Delta \rho_{\gamma}}{\rho_{\gamma}}
~,  \label{eq: eta to rho}
\end{equation}
where we use the relations $n_{\gamma}\propto T^{3}_{\gamma}$ and $\rho_{\gamma} \propto T^{4}_{\gamma}$.

Combining these relations, \citet{Nakama:2014vla} provides a constraint on the density fraction of the injected energy between these epochs as
\begin{equation}
\frac{\Delta\rho_{\gamma}}{\rho_{\gamma}} < 7.71\times 10^{-2} ~,
 \label{eq: eta constraint}
\end{equation}
where they adopted $\eta_{\rm CMB, obs}={(6.11-0.08)\times 10^{-10}}$ \citep{Ade:2013zuv} and $\eta_{\rm BBN,obs} ={(6.19+0.21)\times 10^{-10}}$ \citep{Nollett:2013pwa} as observation values for CMB and BBN, respectively.
Here, we stress that the bound Eq.~(\ref{eq: eta constraint}) is
intrinsically coming from the difference between the baryon-photon
number ratio at the BBN and CMB era, not from the direct measurement of the CMB temperature or the CMB photon number.
In order to obtain the above constraint, we take into account for the negative and positive errors in the CMB and BBN bound, respectively, in order to obtain conservative limit.
Note that, Eq.~(\ref{eq: eta constraint}) can be reread in terms of the CMB temperature by using the relation: $\Delta\rho_{\gamma}/\rho_{\gamma} = 4 \, \Delta T_{\gamma}/T_{\gamma} + O((\Delta T_\gamma / T_\gamma)^2)$, as $\Delta T_{\gamma}/T_{\gamma} < 1.93 \times 10^{-2}$.

The dissipation of the PMFs can be considered to be one of interesting heating sources, and we name the heating mechanism by the dissipation of PMFs {\it magnetic reheating}.
Before the epoch of recombination, PMFs induce the magnetohydrodynamics~(MHD) modes in the photon-baryon plasma and these MHD modes are damped by the viscosity due to the diffusion process of CMB photons.
In this process, the energy of PMFs dissipates into the CMB photons and subsequently, the baryon-photon number ratio changes.
Therefore, we can put a constraint on PMFs from Eq.~\eqref{eq: eta constraint}.
In next section, we give the formulation for the injected energy from the dissipation of PMFs and show a new constraint on PMFs obtained from {\it magnetic reheating} process. The final section is devoted to the conclusion and summary of our results.

%=========================================
\section{Reheating by decaying magnetic fields}
%=========================================

Let us consider the spatially-averaged injected energy into the CMB photons due to the decaying magnetic fields from the BBN era to the CMB distortion era.
The total injected energy is represented as
\begin{equation}
\frac{\Delta\rho_{\gamma}}{\rho_{\gamma}} = \int^{z_{\mu}}_{z_{\rm i}} {\rm d}z\; \left[ - \frac{1}{\rho_{\gamma}(z)} 
\frac{(1+z)^4}{8\pi} \frac{{\rm d}}{{\rm d}z}
\Braket{\left| \bm{b}(z,\bm{x}) \right|^{2}}
\right] ~, \label{eq: uniform reheating}
\end{equation}
where $\rho_{\gamma}(z)$ is the background energy density of photons which scales as $\propto (1+z)^{4}$ and the brackets denote ensemble average.
We also define $\bm{b}(z,\bm{x}) = (1+z)^{-2}\bm{B}(z,\bm{x})$ as the magnetic fields without the adiabatic decay due to the expansion of the universe.
Here, we set $z_{\mu} = 2.0\times 10^{6}$ and $z_{\rm i} = 1.0\times 10^{9}$, which eras correspond to the inefficient of the double Compton scattering and the neutrino decoupling era, respectively~\citep{Hu:1992dc}.

The evolution of magnetic fields due to the MHD damping can be simply expressed in the Fourier space as
\begin{equation}
b_{i}(z,\bm{k}) = \tilde{b}_{i}(\bm{k}) e^{-(k/k_{\rm D}(z))^2} ~,
\end{equation}
where $\tilde{b}_{i}(\bm{k})$ denotes the initial PMFs and $k_{\rm D}(z)$ represents a wave number of the magnetic fields which
is damped at the redshift $z$, caused by the photon viscosity, similar to the Silk damping of the standard primordial plasma.
The damping wave number can be evaluated by the
mode analysis based on the magneto-hydrodynamics~(MHD).
Depending on the MHD modes and the scales, the damping wave number is different~\citep{Jedamzik:1996wp,Subramanian:1997gi}.
Here we consider the case for the fast-magnetosonic mode,
where Alfv\'en and slow-magnetosonic modes are damped in the photon diffusion limit in
which the mode wavelength $k^{-1}$ is much larger than the photon mean free path.
In such a case, the damping scale is similar to the Silk damping
scales and we adopt~\citep{Jedamzik:1999bm}
\begin{equation}
 k_{\rm D}(z) = 7.44 \times 10^{-6} (1+z)^{3/2} ~ {\rm Mpc}^{-1} ~, \label{eq: diff k}
\end{equation}
in the radiation dominated epoch.
We will discuss the other cases later.

We assume that the initial PMFs are statistically homogeneous and isotropic Gaussian random fields.
The power spectrum of such fields can be written as
\begin{equation}
\braket{\tilde{b}_{i}(\bm{k})\tilde{b}^{*}_{j}(\bm{k'})} = (2\pi)^{3}\delta^{3}_{\rm D}(\bm{k} - \bm{k'}) \frac{\delta_{ij} - \hat{k}_{i}\hat{k}_{j}}{2} \frac{2\pi^{2}}{k^{3}}\mathcal{P}_{B}(k) ~,
\end{equation}
and then the bracket of the right-hand side in Eq.~(\ref{eq: uniform reheating}) can be expressed as
\begin{equation}
\Braket{\left| \bm{b}(z,\bm{x}) \right|^{2}}
=
\int {\rm d}\ln k \,\,
\mathcal{P}_{B}(k)\,
e^{-2 \left( \frac{k}{k_{\rm D}(z)} \right)^{2}} ~. \label{eq: braket bb}
\end{equation}
Therefore, Eq.~\eqref{eq: eta constraint} allows us to obtain the constraint on the power spectrum of magnetic fields.
In our analysis, we employ two types of the power spectrum, one is
a delta-function type in logarithmic scale given by
\begin{equation}
\mathcal{P}_{B}(\ln k) = \mathcal{B}_{\rm delta}^{2} \,\delta_{\rm D}(\ln (k/k_{\rm p})) ~, \label{eq: delta type}
\end{equation}
where $\mathcal{B_{\rm delta}}$ corresponds to the amplitude at the peak wave number $k_{\rm p}$,
and another is a power-law type given by
\begin{equation}
\mathcal{P}_{B}(k) = \mathcal{B}^{2} \left( \frac{k}{k_{\rm n}} \right)^{n_{B}+3} ~, \label{eq: power law type}
\end{equation}
where $\mathcal B$ is the amplitude at the normalization wave number
$k_{\rm n}$.
In this Letter, we set the normalization scale as $k_{\rm n} = 1~{\rm Mpc}^{-1}$.

\begin{figure}
\begin{center}
\includegraphics[width=0.49\textwidth]{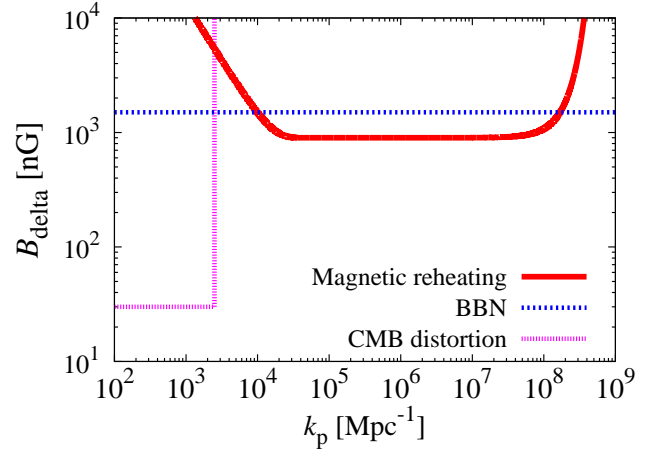}
\end{center}
\caption{%
The upper bound on the amplitude of the delta-function type for the power spectrum $\mathcal{B}_{\rm delta}$ as a function of $k_{\rm p}$ from the magnetic reheating.
The limits from the BBN \citep{2012PhRvD..86f3003K} and CMB distortions \citep{Jedamzik:1999bm} in blue and magenta lines, respectively, are shown in the same plot.
}
\label{fig: limit B delta}
\end{figure}

First we focus on the delta function type for the power spectrum.
By substituting Eq.~(\ref{eq: delta type}) into Eq.~(\ref{eq: uniform reheating}) with Eq.~(\ref{eq: braket bb}),
we obtain the injected energy fraction to the CMB energy by the decaying PMFs as
\begin{equation}
\label{eq:deltaform}
\frac{\Delta\rho_{\gamma}}{\rho_{\gamma}}
=
\frac{\mathcal{B}_{\rm delta}^{2}}{8\pi \rho_{\gamma, 0}} \tilde{C}(k_{\rm p}) ~,
\end{equation}
where $\tilde{C}(k_{\rm p})$ is the function of $k_{\rm p}$ and which the explicit form is given as
\begin{equation}
\tilde{C}(k) \equiv \exp{\left[ -2\frac{k^{2}}{k^{2}_{\rm D}(z_{\rm i})}\right]} - \exp{\left[ -2\frac{k^{2}}{k^{2}_{\rm D}(z_{\mu})}\right]} ~.
\end{equation}
When $k_{\rm p}$ is much larger than $k_{\rm D} (z_{\rm i})$ ($> k_{\rm D} (z_\mu)$),
the energy injection due to the decay of the PMFs becomes efficient before the BBN era.
On the other hand, when $k_{\rm p}$ is much smaller than $k_{\rm D} (z_\mu)$,
the energy injection from the PMFs occurs during the CMB distortion era.
We focus on the case where the energy injection from the decaying of the PMFs
is efficient after the BBN era and before the CMB distortion era,
that is, the case where
$k_{\rm p}$ is set between the two decaying scales $k_{\rm D}(z_{\rm i})$ and $k_{\rm D}(z_{\mu})$.
For such a case, we can approximately take $\tilde{C}(k_{\rm p}) \to 1$~(see~\citealt{Miyamoto:2013oua}).
Combining Eq.~\eqref{eq: eta constraint} and Eq.~\eqref{eq:deltaform} yields a constraint on PMFs in the $k_{\rm p}$-${\mathcal B}_{\rm delta}$ plane.
We present our new result obtained from magnetic reheating as a red line in Fig.~\ref{fig: limit B delta}.
%===
%===
For comparison, we plot limits from the BBN and CMB distortions in magenta and blue lines, respectively.
We can see that our magnetic reheating constraint gives a tight limit on the PMFs on small sales from $k_{\rm p} =10^4~\rm Mpc^{-1}$ to $k_{\rm p} = 10^8~{\rm Mpc}^{-1}$.

\begin{figure}
\begin{center}
\includegraphics[width=0.49\textwidth]{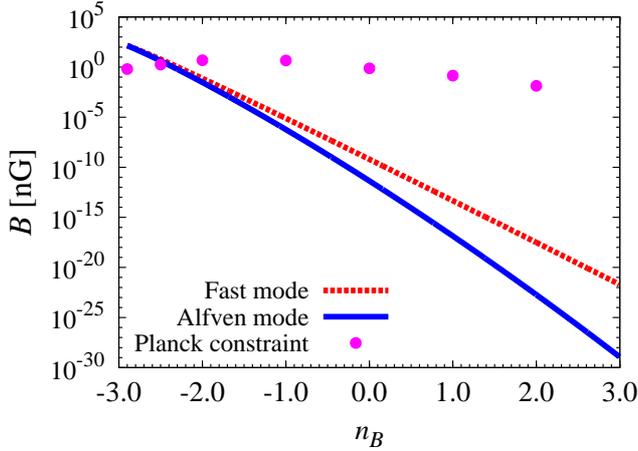}
\end{center}
\caption{%
The upper bound on the amplitude of the power-law type for the power spectrum $\mathcal{B}$ as the function of $n_{B}$ from the magnetic reheating of the fast-magnetosonic mode (red) and Alfv\'en-magnetosonic mode (blue).
The constraint from the Planck \citep{2016A&A...594A..19P} is also shown.
Note that $n_{B} = -3.0$ corresponds to the scale-invariant power spectrum.
}
\label{fig: limit B power}
\end{figure}

Next we consider the power-law type of the power spectrum defined in Eq.~(\ref{eq: power law type}).
By using Eq.~(\ref{eq: uniform reheating}) with Eq.~(\ref{eq: braket bb}), we obtain 
\begin{equation}
\frac{\Delta\rho_{\gamma}}{\rho_{\gamma}}
=
\frac{\mathcal{B}^{2}}{8\pi \rho_{\gamma, 0}}
\frac{\Gamma\left( \frac{n_{B}+3}{2}\right)}{2^{(n_{B}+5)/2}}
\left[ \left( \frac{k_{\rm D}(z_{\rm i})}{k_{\rm n}}\right)^{n_{B}+3} - 
\left( \frac{k_{\rm D}(z_{\mu})}{k_{\rm n}}\right)^{n_{B}+3} 
\right] ~.
\label{eq:delta_rho/rho}
\end{equation}
Here, we focus on the magnetic reheating due to the fast-magnetosonic mode given in Eq.~(\ref{eq: diff k}).
In Fig.~\ref{fig: limit B power}, we plot a limit for the amplitude of the power-law type as a function of the spectral tilt $n_{B}$ in a red line.
%===
%===
We find that blue-tilted spectrum is strongly constrained compared with the scale-invariant spectrum ($n_{B} = -3.0$).
It is shown that the causal mechanism of PMFs generation predicts only a
blue spectrum with $n_B \ge 2$~\citep{2003JCAP...11..010D}.
Therefore, the constraint on the blue-tilted spectrum is suggestive for such a causal mechanism.

So far we have considered the case that the Alfv\'en and slow-magnetosonic
modes are damped in the photon diffusion limit. However, depending on
the magnetic field strength, the energy damping of these modes are
insufficient in the photon diffusion limit because the magnetic field
cannot accelerate the fluid efficiently.
When it happens, the magnetic field can survive even below smaller
scale than $k^{-1}_{\rm D}(z)$ given in Eq.~\eqref{eq: diff k}.
These modes are damped later
when the scale of the modes are smaller than the photon free-streaming scale. 
This damping scale is roughly given by $k_{\rm D}^{\rm A}(z) \sim k_{\rm D}(z)/V_{\rm A}(z)$
in the radiation dominated epoch. Here $V_{\rm A}(z)$ is the
Alfv\'en velocity, $V_{\rm A}(z) = B_{\lambda}(z)/\sqrt{16 \pi
\rho_{\gamma,0}/3}$,
where we obtain the magnetic field $B_\lambda(z)$ by integrating the power
spectrum over $k$
with a Gaussian window function with a scale $\lambda(z)$
corresponding to $k_{\rm D}^{\rm A}(z)$~\citep{2002PhRvD..65l3004M}.
Therefore, $k_{\rm D}^{\rm A}(z)$ can be represented as
\begin{equation}
\frac{k_{\rm D}^{\rm A} (z)}{k_{\rm n}}
\sim
2\pi
\left[
\frac{ 3.5 \times 10^{5}}{\Gamma\left(\frac{n_{B}+5}{2}\right)}
\left(\frac{\cal B}{1~\rm nG}\right)^{-2}
 \left( \frac{k_{\rm D}(z)}{k_{\rm n}}\right)^{2}
\right]^{1/(n_B + 5)}.
 \label{eq:kda}
\end{equation}
Replacing $k_{\rm D}(z)$ to $k_{\rm D}^{\rm A}(z)$ in Eq.~\eqref{eq:delta_rho/rho},
we can evaluate the injected energy fraction for the Alfv\'en and
slow-magnetosonic modes in the photon free-streaming limit
and plot the constraint on primordial magnetic fields as a blue line in Fig.~\ref{fig: limit B power}.
We find that the magnetic reheating due to the Alfv\'en and
slow-magnetosonic modes put a stronger constraint on the amplitude of PMFs than the fast-magnetosonic mode.
The obtained constraint can be roughly fitted as $\log({\cal B}/{\rm nG}) \lesssim -11-6n_B$
(for $n_B \gtrsim 0$).
This constraint is tighter than other CMB constraints.

%=========================================
\section{conclusion}
%=========================================

Before the CMB distortion era, the additional energy injection
heats the CMB photons and due to the double Compton scattering, the number of the CMB photons increases.
Therefore, the baryon-photon number ratio decreases by the reheating.
By comparing the values of the baryon-photon number ratio between
two epochs, we can determine the allowed energy injection or allowed
changes of the CMB temperature
between them.
In this Letter, we assume the reheating source as the decaying PMFs due to the photon viscosity, named the magnetic reheating.
Since the baryon-photon number ratio is well constrained by the current
cosmological observations such as the BBN and CMB, the magnetic
reheating provides us a significant constraint on small-scale PMFs
which we cannot access through the observations of CMB anisotropies and distortions.
From the state-of-the-art observations, we put the new constraint on the
amplitude of small-scale PMFs as $B_{0} \lesssim 1\; \mu{\rm G}$ in the
range $10^{4} \; h{\rm Mpc}^{-1} < k < 10^{8} \; h{\rm Mpc}^{-1}$.

We also apply the magnetic reheating constraint to PMFs with
a power-law spectrum.
Since the dissipation scale depends on the MHD modes,
we consider two damping scales of
not only the fast-magnetosonic mode but also the slow-magnetosonic and
Alfv\'en modes.
We find that for blue-tiled spectrum
the slow-magnetosonic and Alfv\'en modes can set stronger constraint, compared with
the constraint from the fast-magnetosonic mode.
In general, because the damping scale of the slow-magnetosonic mode and Alfv\'en
mode is smaller than that of the fast-magnetosonic mode
the slow-magnetosonic and Alfv\'en modes are sensitive to the PMFs on
small scales and can set a stronger constraint for blue-tiled spectrum 
rather than the fast-magnetosonic mode.
The obtained constraint can be roughly fitted as $\log({\cal B}/{\rm nG}) \lesssim -11-6n_B$
(for $n_B \gtrsim 0$) where $\cal B$ is the PMF amplitude normalized at 1~comoving scale.
This constraint is tighter than other CMB constraints in the case of the blue-tilted spectrum.

Although we consider two damping scales separately in this Letter,
in reality, the energy dissipation of magnetic fields happens through
the combination of these two damping scale channels.
However it is not so trivial how much percentage of the magnetic field
energy is dissipated through the fast-, slow-magnetosonic or Alfv\'en
modes.
When the dissipated energy is in equipartition between these modes, our
constraint might be relaxed by a few factor.
To investigate the details, we will address this issues through the analysis
in higher order cosmological perturbations.

Among many mechanisms to generate PMFs in the early universe, 
the causal mechanism can generate only blue power spectrum~\citep{2003JCAP...11..010D}.
Therefore, strong constraints for the blue-tiled PMFs obtained in this Letter should be a powerful tool to investigate a successful causal mechanism to generate PMFs in the early universe.

%=========================================
\section*{Acknowledgements}
This work is supported in part by a Grant-in-Aid for JSPS Research Fellow Number 17J10553~(S.S.),
JSPS KAKENHI Grant Number 15K17646~(H.T.), 17H01110~(H.T.) and 15K17659~(S.Y.),
and MEXT KAKENHI Grant Number 16H01103~(S.Y.).

\bibliographystyle{mnras}
%---------- Bibtex ----------%
\bibliography{ref}
\end{document}